# Effect of charge ordering on the electrical properties and magnetoresistance of manganites

Mahrous R Ahmed

Department of Physics, Faculty of Science, Sohag University, Sohag 82534, Egypt.

Correspondence e-mail: mahrous.r.ahmed@science.sohag.edu.eg

## Abstract

The Monte Carlo Ferromagnetic Ising model was used to study the electrical properties of manganese oxides due to the charge ordering phase occurring at doping, $x = 0.5$. The half-doped manganites have an insulator antiferromagnetic ground state. We calculated the internal energy, specific heat, resistivity and the magneto-resistance, MR, with parallel and anti-parallel applied magnetic fields. Our simulation reveals that the resistivity decreases exponentially and the electric current increases with increasing temperature according the free charge increase, to transport from an insulator to conductor phase. The magnetoresistance has negative small values with parallel magnetic field but has positive high values with unti-parallel magnetic field. The obtained semiconductor-metal transition behavior candidates the half-doped manganites to be very good semiconductors diode junctions.

# Introduction:

Perovskites compounds have attracted much interest from more than fifty years because of their various properties that are too important to apply in the recent electronic devices such as memory discs and smart devices[1-7]. Perovskites compounds such as manganites, namely, $RE_{1-x}A_xMnO_3$ (RE = rare earth, A = Ca, Sr, Ba and Pb) have attracted considerable interests according to the discovery of the colossal magnetoresistance (CMR) [8-11] as well as many other exciting phases such as the intrinsic ferromagnetism, orbital ordering[12-16], charge–orbital[17] and charge- ordering phases[18]. Charge-ordering (CO) phase occurs mostly in strongly correlated materials such as mixed-valent transition metal oxides especially manganites compounds with half-doping, x=0.5[19]. In this phase, charges are localized on different sites leading to a high insulating and an ordered superlattices, where the positive charge, $Mn^{4+}$, alternate with the negative, $Mn^{3+}$, and are ordered antiferromagnetically along the lattice.

There have been many debates about how is the charge ordering patterns [19], some experimental researches showed that charge-ordering phase has stripes patterns [20-24], but other many researches [25-32] assured that it has a checkerboard configuration. The checkerboard charge ordering and the orbital ordering are the result of the Jahn-Teller interaction of the $e_g$ orbitals with the lattice[27]. Subias et al noted that the low temperature phase is described as a checkerboard ordering of two types of Mn sites with different local geometrical structures[28]. However, theoretical works in this interesting research obtained results confirmed that the charge-ordering phase has first order phase with 3d antiferromagnetic ground state [33].

The charge-ordered is insulating state, which transforms to a metallic ferromagnetic state on the application of a magnetic field. It has been experimentally obtained [34-36] that resistivity decreases rapidly in first order transition from insulator to metal-like at charge-order transition temperature, $T_{CO}$, with different applied magnetic field. RAO et. al. [37] theoretically showed the same results that the half-doped manganites have resistivity changing from insulator to metal phase in first order manner as function of the temperature and as function of an applied magnetic field using double exchange model and other theoretical methods.

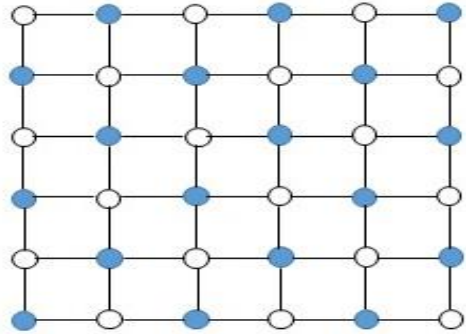

**Fig. (1):** (Colour figure online)**The charge ordering, CO, phase, which occurs in half-doped manganites. The empty circles are Mn$^{4+}$ representing the positive charge and filled circles are Mn$^{3+}$ representing the negative charge**

## The model:

Charge-ordering phase occurs in Half-doped manganites showing insulator behavior at low temperature. In the charge-ordering ground state configuration both Mn$^{3+}$, the negative charge, and Mn$^{4+}$, the positive charge, order antiferromagnetically in the lattice through the three dimensions, as if it is checkerboard ordering, see Fig. (1). Now, we have two states interacting electrically to each other, Mn$^{3+}$ which has an electron occupying the $e_g$ orbital, $d_{3z3-r2}$ and $Mn^{4+}$ which has missed an electron from the same orbit to create a hole instead.

The simplest model we can use to describe the interaction between these our two states is Ising model. Ising model described earlier [38] is usually used to study statistically the interaction between two nearest neighbouring states in 2d or 3d. We use in this work the three-dimensional Ising model with two-states consists of *N* sites ($N=L^3$) where L is the number of sites for every dimension of the lattice, here *L=10*. Because we are interested in the simulation of the 3d Ising model for different temperatures and variable applied magnetic field, the Hamiltonian can be expressed in the form,

$$U = -\frac{J}{2}\sum_{<i,j>}^{N} S_i S_j - H \sum_i^N S_i \qquad (1).$$

Here, <i,j> means that the sum is over the nearest neighbor pairs of atoms and $S_i$ and $S_j$ are the charges of $Mn^{3+}$ and $Mn^{4+}$ in i and j sites. Furthermore, *J* is the exchange energy between each two nearest neighbors. The applied magnetic field is either parallel to the direction of the flow of the positive charges, $Mn^{+4}$, i.e., the direction of the electric field or antiparallel to the positive charges, see Fig. (2). Here *1/2* is to avoid the double counting along the lattice.

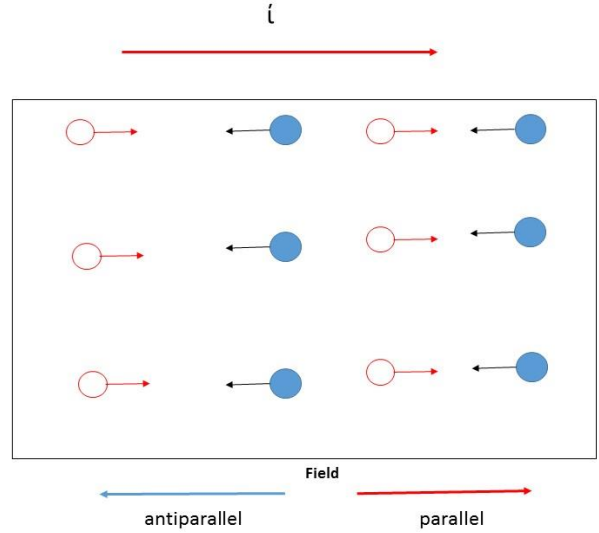

**Fig. (2):** (Colour figure online) The Electric field is applied in either parallel or unparallel to the direction of the flow of the positive charges, current i,. The empty circles are positive charges, $Mn^{4+}$, and the filled circles are the negative charges, $Mn^{3+}$.

Monte Carlo (MC) method and Metropolis algorithm are used to simulate our system with averaging performed $10^6$ MC steps. The system is antiferromagnetically ordered along the three direction, so that the symmetry is broken at the charge ordering transition temperature, $T_{CO}$. We obtained Results by both cooling down from a high-temperature random configuration as discussed by Banavar et. al. [39] and heating up from the antiferromagnetic, AF, ground state. The results of both cooling and heating are consistent.

## Results and Discussion

Studying the electrical and magnetoresistance properties for the half-doped manganites is very important because of its interesting application in the electronical devices. As we noted above we use here the Ising model to study the interaction between charges ordering in the lattice of the half-doped manganites and to show how they contribute to the charge current along the lattice with the change of the temperature and the applied magnetic field.

The internal energy, E, of our system, as a function of $K_BT/J$ at both parallel and anti-parallel applied magnetic field, H, is shown in Fig. (3). The internal energy is always decreasing with increasing of the temperature in a weak first order transition from parallel magnetic field to strong first order transition at antiparallel magnetic field. It could be attributed to the increasing of the number of the free charges those iterate through the lattice contributing to the electric current. At zero-temperature the internal energy equals zero for all values of the magnetic fields, but at high temperature, $T_H$, the internal energy obey the following formula:

$$E(T_H) = -\frac{1}{2}\left(\frac{6J}{2}\right) \qquad (2)$$

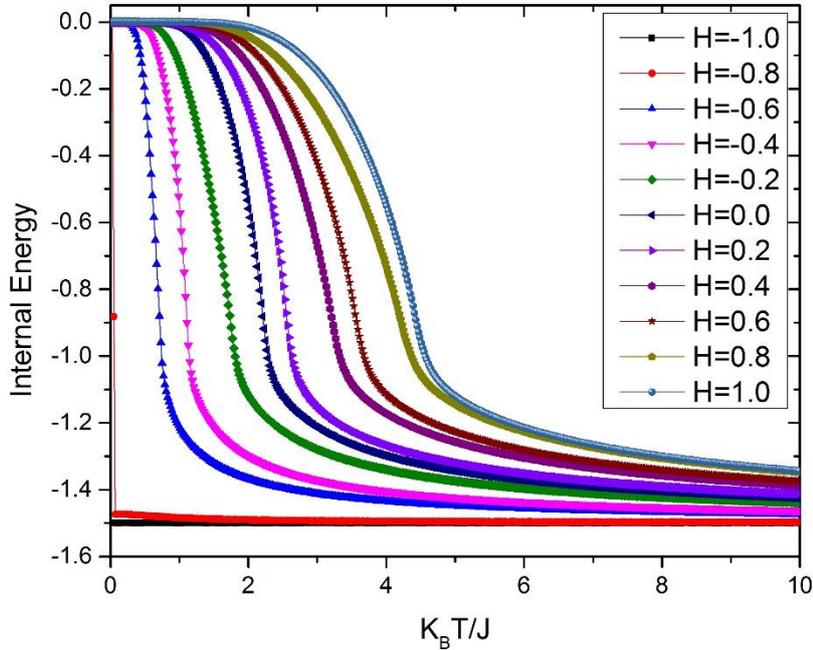

Fig. (3): (Colour figure online) Internal energy, E, as a Function of $K_BT/J$ at both parallel and anti-parallel applied magnetic fields for 3-d Two-states Ising model.

At T→ ∞, E → $\frac{6J}{4}$ = 1.5 energy unit. That means that only the quarter of charges are ordered at high temperature.

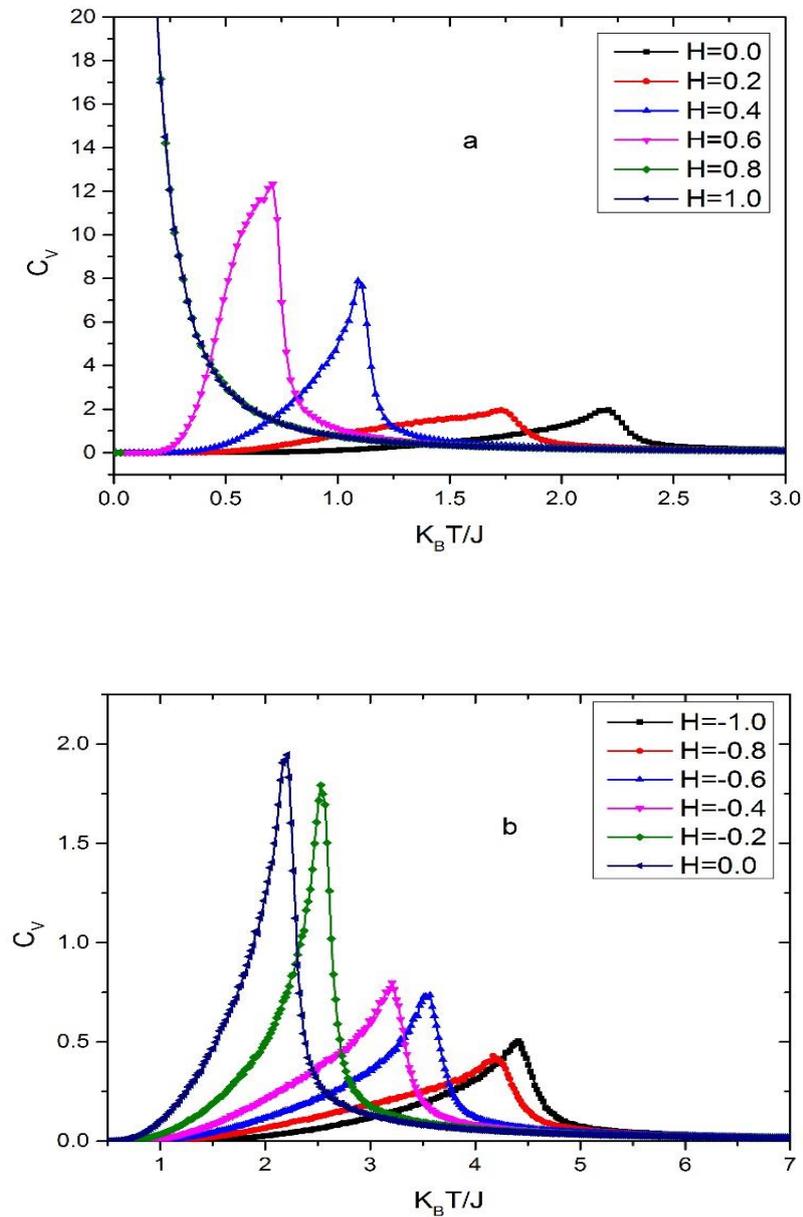

Fig. (4): (Colour figure online) Specific heat, $C_V$, as a function of $K_BT/J$ at both (a) parallel and (b) anti-parallel applied magnetic fields for 3-d two-states Ising model.

The dependence of the specific heat, $C_V$, of the temperature ($K_BT/J$) in half-doped manganite is shown in Fig. (4) for both parallel, Fig. 4(a), and anti-parallel, Fig. 4(b), different applied magnetic fields.

It is noticed from Fig. 4(a) that the specific heat has very large maximum value at high parallel fields, i.e., at $H =1.0$ and $0.8$ compared with the rest of the applied magnetic field range, but the maximum value decreases with the decreasing H until the end of the anti-parallel range as shown in Fig. 4(b). Also one can see from the specific heat results that the transition is a second order transition along the range of $H$ except at $H=1.0$ and $0.8$ it has a first order transition.

From $C_V$ results, we can obtain the charge ordering transition temperatures, $T_{CO}$, of our system. Fig. (5) Shows the effect of the applied magnetic fields on the charge ordering transition temperatures $T_{CO}$. We could deduce from the results that $T_{CO}$ decrease with increasing $H$ as follow:

$$T_{CO} \; \alpha \; H^{\gamma} \qquad (3).$$

After fitting the $H$-$T_{CO}$ results we got, $\gamma = -2.23 \pm 0.06$ and the intercept $= 2.172 \pm 0.09$.

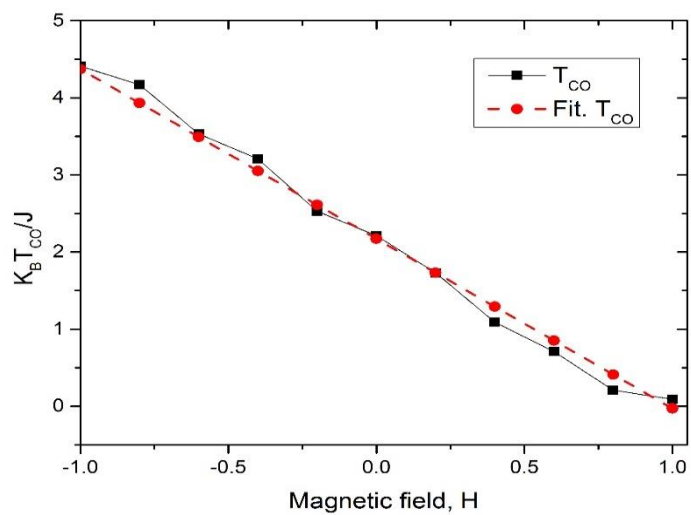

**Fig. (5):** (Colour figure online)**The charge ordering transition temperature, $T_{CO}$, as function of the applied magnetic field.**

When the temperature increase the localized charges of $Mn^{3+}$ free and contribute to the current and become positive charges, $Mn^{4+}$, then the resistivity decreases dramatically with increasing the temperature as shown in Fig. (6). The resistivity decreases in first order pattern and shows semiconductor-like behavior for all values of the applied magnetic fields except for $H = 0.8$ and $1.0$ the resistivity decreases very slowly as it becomes too small compared to its values along the range $-1.0 \leq H \leq 0.6$. The resistivity is calculated from the charge obtained from our simulation as follow:

$$\rho = k/q \qquad (4).$$

$k$ is a constant, $k = vtd$, where we considered the voltage difference is a constant and d is the sample thickness. We put the time, $t = 1$. but at high $T$ the resistivity tends to be too small along the range of $H$.

At low $T$, $\rho$ has values of order $10^4$ which agree with the theoretical[40] and experimental published results for $Sm_{1/2}Ca_{1/2}MnO_3$[41,42], $Pr_{1-x}Ca_xMnO_3$ [43,44], $La_{0.5}Ca_{0.5}MnO_3$[45] and $Dy_{0.5}(Sr_{1-x}Ca_x)_{0.5}MnO_3$[46]

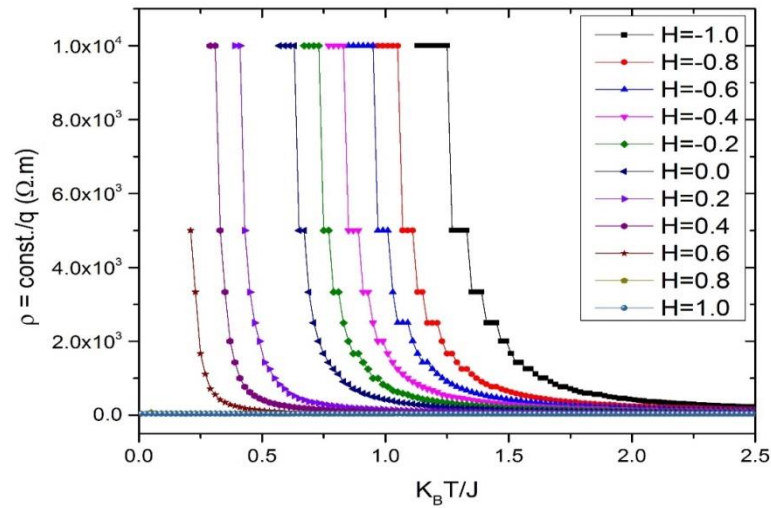

**Fig. (6):** (Colour figure online)**The resistivity, ρ, as a function of $K_BT/J$ at both parallel and anti-parallel applied magnetic fields.**

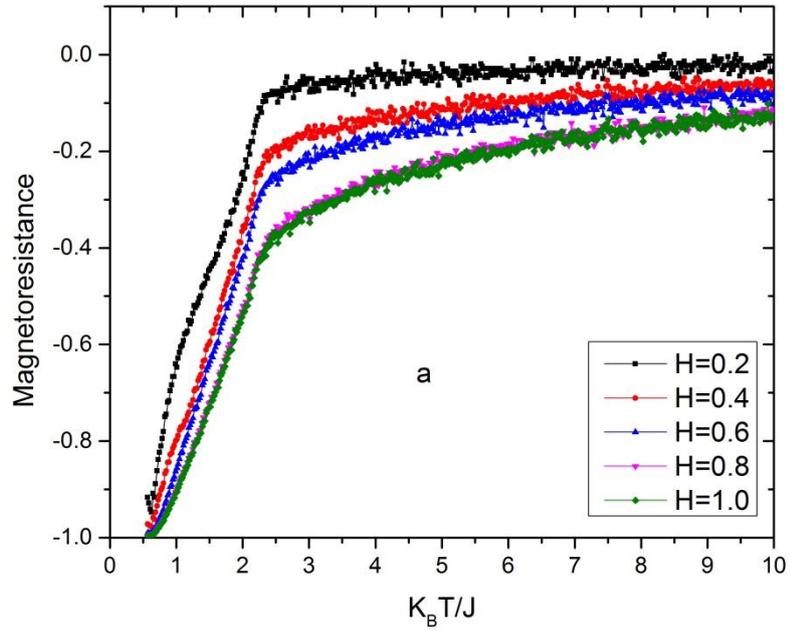

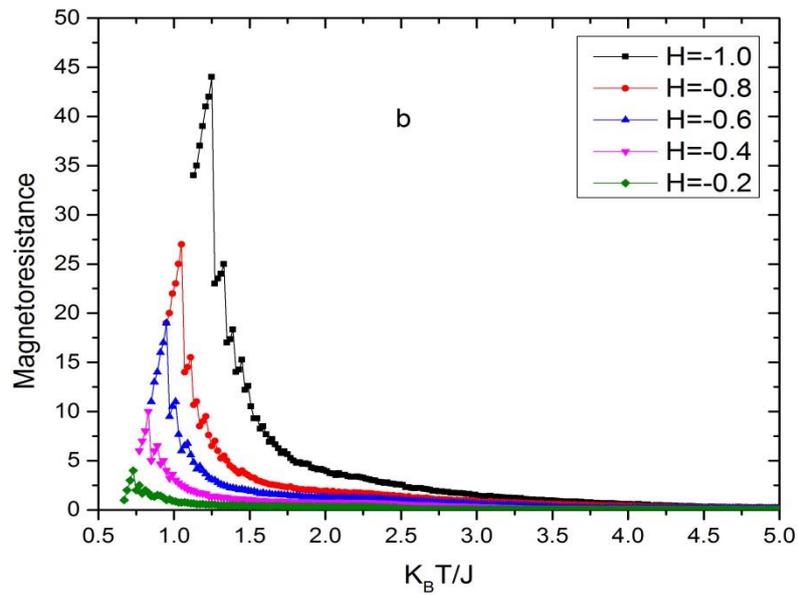

**Fig.(7)** (Colour figure online)**Magnetoresistance as function of Temperature with parallel magnetic field (a) and unti-parallel magnetic field (b).**

Magnetoresistances as function of temperature at different values of applied parallel and unti-parallel magnetic fields are shown in Fig 7(a) and fig. 7(b) respectively. The magnetoresistance, MR, could be calculated from the following equation:

$$MR = \frac{\rho(H) - \rho(0)}{\rho(0)} \qquad (5),$$

where $\rho(H)$ is the resistivity with non-zero magnetic field and $\rho(0)$ is the resistivity with zero magnetic field. It is found from Fig. 7(a) that when we apply the magnetic field in parallel to the charge current the MR has small and negative values. Besides, It increases with increasing temperature in first order manner until $T_{CO}$. In addition, after $T_{CO}$ MR becomes stable gradually with temperature, while *MR* decreases with increasing the applied magnetic field. At parallel magnetic field, the resistivity becomes smaller than the resistivity at $H = 0$, so that MR is negative.

When we apply unti-parallel magnetic field as shown in fig. 7(b) *MR* increases linearly with T until it reaches $T_{CO}$, whereas after $T_{CO}$, it decreases promptly. *MR* has positive and higher values than those of MR at parallel magnetic field. The resistivity, in case of unti-parallel *H* is found to be higher than that of $H = 0$, so that *MR* is positive.

*MR* has very small and negative values when the magnetic field is applied in parallel, while it has considerable and positive values when the magnetic fields is applied in unti-parallel to the charge current. This is the semiconductor-like behavior where the resistivity has high values at low temperature, decreases exponentially with the temperature and does not obey the Ohm's law as shown in Fig. (6).

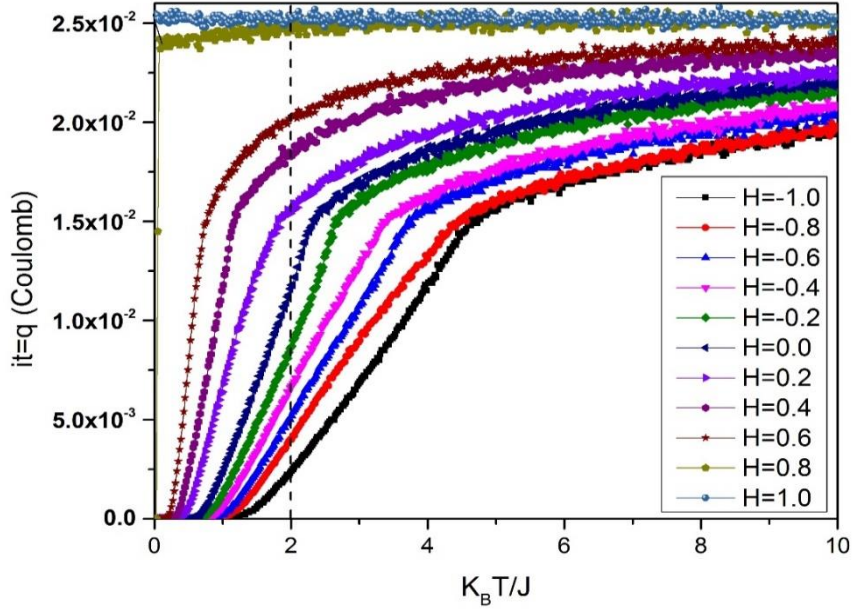

**Fig. (8):** (Colour figure online)**The electric current, i, as a function of $K_BT/J$ at both parallel and anti-parallel applied magnetic fields.**

Fig. (8) shows that the charge current in half-doped manganites increases in second order transition along the whole range of H, except for H=0.8 and 1.0 it increases rapidly in a first order manner where the magnetic field strength overwhelm the entire charge bonds to orient them easily in order to contribute to the electric current. The current varies from zero to $25 \times 10^{-3}$ of the current unit. The current could be calculated from our simulation as following:

$$I = q/t \qquad (6).$$

We put t = 1. It is noticed that the current increases quickly with increasing of the applied magnetic field until the transition temperature, $T_{CO}$, after that the current saturates gradually before it stabilize at very high temperature.

Figure (9) shows the relation between the charge current and the applied magnetic field at a constant temperature, $K_BT/J = 2$, which is specified by a vertical dashed line in Fig. (8). We can deduce how the current depend on $H$, $I\alpha H^\beta$, by fitting the I-H results as shown in Fig. (9). We obtained β = 0.0128.

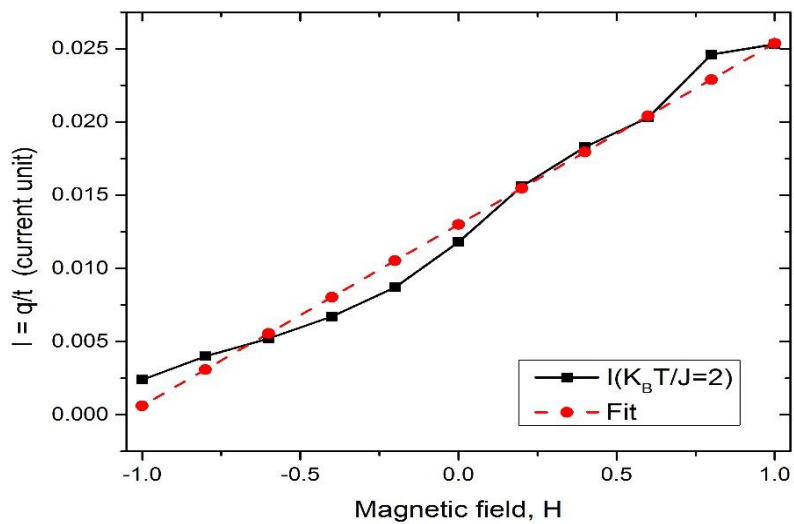

**Fig. (9):** (Colour figure online) **The current as function of magnetic field at $K_BT/J=2$. The red line is the fitting of the I-H results.**

We can interpret our results in light of the existing of a specific strong competition between the electron-lattice and the double exchange interaction. The strong electron-lattice interaction prefers the localization of charge carriers, in the same time the double exchange ferromagnetism requires the active hopping of charge carriers. Therefore, it is the competition between these two interactions that is mainly responsible for the charge transport properties and the settling of CO states.

The role of the applied magnetic field is to align spins along its direction and thus results in the increase of the effective strength of double exchange interaction, which in turn improves the FM coupling between neighboring Mn spins with the mobility of charge carriers, increasing the entire size of the FM domains and reducing that of the insulating CO domains. The metallic FM domains are spread in the whole sample, taking us to a dramatic decrease in the resistivity[47,48].

# Conclusion

Many manganites have charge ordering phase at half-doping which is very important phase for a lot of artificial application in electronic devices. In this work we used two states 3-d Ising model to study the electrical properties and magnetoresistance with and without applied magnetic field for the manganites with the charge ordering phase.

Our results showed semiconductors-metals transition at $T_{CO}$ with first ordering manner at most of applied magnetic fields. The applied magnetic field effects on the studied properties are appeared clearly. The charge of the current increases with increasing the magnetic field with an order equal to 0.0128, while the $T_{CO}$ decreases with increasing of the magnetic field with an order equal to -2.23.

The magnetoresistance is negative and very small when the magnetic field is applied in parallel to the charge current but it is high and positive with anti- parallel magnetic field. MR results confirm that half-doped manganites have semiconductors like behavior.

We interpreted our results in light of the specific strong competition between the electron-lattice and the double exchange interaction which reduce the insulating CO domains and increase the entire size of the FM domains that spread in the whole sample, taking us to a dramatic decrease in the resistivity.

# Acknowledgments

# References

**[1]** R. von Helmolt, J. Wecker, B. Holzapfel, L. Schultz, and K. Samwer Phys. Rev. Lett. **71**, 2331 (1993).

[2] S. Jin, T. H. Tiefel, M. McCormack, R. A. Fastnacht, R. Ramesh and L. H. Chen, Science 264, 413 (1994).

[3] H. Kuwahara, Y. Tomioka, A. Asamitsu, Y. Moritomo and Y. Tokura, Science 270, 961 (1995).

[4] T. Obata, T. Manako, Y. Shimakawa and Y. Kubo, Appl. Phys. Lett. 74, 290 (1999).

[5] X. W. Li, Y. Lu, G. Q. Gong, G. Xiao, A. Gupta, P. Lecoeur, J. Z. Sun, Y. Y. Wang and V. P. Dravid, J. Appl. Phys. 81, 5509 (1997).

[6] G. Prinz, Phys. Today 48, 58 (1995).

[7] M. Rajeswari, C. H. Chen, A. Goyal, C. Kwon, M. C. Robson, R. Ramesh, T. Venkatesan and S. Lakeou, Appl. Phys. Lett. 68, 3555 (1996).

[8] Liu Yu-Kuai, Yin Yue-Wei and Li Xiao-Guang, Chin.Phys.B 22, 087502 (2013).

[9] S. Jin, T. H. Tiefel, M. McCormack, R. A. Fastnacht, R. Ramesh and L. H. Chen, Science 264, 413 (1994).

[10] R. Mahesh, R. Mahendiran, A. K. Raychaudhuri and C. N. R. Rao, J. Solid State Chem. 114, 297 (1995).

[11] L. Sheng, D. Y. Xing, D. N. Sheng, and C. S. Ting, Phys. Rev. Lett. **79**, 1710 (1997).

[12] K. I. Kugel and D. I. Khomskii , Sov. Phys.–Usp. 25, 231(1982).

[13] Y. Tokura and N. Nagaosa, Science 288, 462(2000).

[14] J. Van der Brink, G. Khaliullin and D. Khomskii, Colossal Magnetoresistance Manganites, KLUWER ACADEMIC PUBLISHERS DORDRECHTI BOSTON I LONDON, ISBN 978-90-48\-6527-8, DOl 10.1007/978-94-0 J 5-1244-2 (2004).

[15]  M. R. Ahmed and G. A. Gehring, J. Phys. A: Math. Gen. 38, 4047 (2005).

[16] Mahrous R. Ahmed and G. A. Gehring, Phys. Rev. B 74, 014420 (2006).

[17] Y. Tokura and N. Nagaosa, Science 288, 462 (2000).

[18] C. N. R. Rao, Science. 276, 911(1997).

[19] J.P. Attfield, Solid State Sciences 8, 861 (2006).

**[20]** A. J. Millis, P. B. Littlewood, and B. I. Shraiman

Phys. Rev. Lett. **74**, 5144 (1995).

**[21]** Y. Tokura, H. Kuwahara, Y. Moritomo, Y. Tomioka, and A. Asamitsu, Phys. Rev. Lett. **76**, 3184 (1996).

[22] J. Geck, D. Bruns, C. Hess, R. Klingeler, P. Reutler, M. v. Zimmermann, S.-W. Cheong, and B. B chner, Phys. Rev. B 66, 184407 (2002).

[23] J. C. Loudon, N. D. Mathur, and P. A. Midgley, Nature 420, 797 (2002).

[24] F. M. Woodward, J. W. Lynn, M. B. Stone, R. Mahendiran, P. Schi er, J. F. Mitchell, D. N. Argyriou, and L. C. Chapon, Field-induced

avalanche to the ferromagnetic state in the phase-separated ground state of manganites, Phys. Rev. B 70, 174433 (2004).

[25] H. Aliaga, D. Magnoux, A. Moreo, D. Poilblanc, S. Yunoki, E. Dagotto, PHYSICAL REVIEW B 68, 104405 (2003).

[26] WANG Hai-Long, TIAN Guang-Shan, and LIN Hai-Qing, Commun. Theor. Phys. 43, 179 (2005).

[27] P Schlottmann, Journal of Physics: Conference Series 200, 012177 (2010).

[28] G. Subías, J. García, P. Beran, M. Nevřiva, M. C. Sánchez, and J. L. García-Muñoz, Phys. Rev. B 73, 205107 (2006.)

[29] H. Aliaga, D. Magnoux, A. Moreo, D. Poilblanc, S. Yunoki, and E. Dagotto, Phys. Rev. B 68, 104405 (2003).

[30] Gianluca Giovannetti, Sanjeev Kumar, Jeroen van den Brink, and Silvia Picozzi5, PRL 103, 037601 (2009).

[31] P. G. Radaelli, D. E. Cox, M. Marezio and S-W. Cheong, Phys. Rev. B 55, 3015 (1997).

[32] James C. Loudon, Neil D. Mathur & Paul A. Midgley, NATURE VOL 420, 19 (2002).

 [33] O. Cépas, H. R. Krishnamurthy, and T. V. Ramakrishnan, Phys. Rev. B **73**, 035218 (2006)**.**

[34] H. Aliaga, D. Magnoux, A. Moreo, D. Poilblanc, S. Yunoki, and E. Dagotto, Phys. Rev. B 68, 104405 (2003).

[35] J. Geck, D. Bruns, C. Hess, R. Klingeler, P. Reutler, M. v. Zimmermann, S.-W. Cheong, and B. Büchner, Phys. Rev. B 66, 184407 (200).

[36] R. Mahendiran, S. K. Tiwary, A. K. Raychaudhuri, and T. V. Ramakrishnan, Phys. Rev. B53, 3348 (1996).

[37] C.N.R. Rao and A.K. Raychaudhuri, Colossal Magnetoresistance, Charge Ordering and Related Properties of Manganese Oxides, World Scientific, . pp. 1-42(1998).
DOI: http://dx.doi.org/10.1142/9789812816795_0001.

[38] E. Ising, Z. Phys. 21, 613 (1925).

[39] J. R. Banavar, G. S. Grest, and D. Jasnow, Phys. Rev. B 25, 4639 (1982).

[40] H. Aliaga, D. Magnoux, A. Moreo, D. Poilblanc, S. Yunoki, and E. Dagotto, Phys. Rev. B 68, 104405 ( 2003).

[41] J. L. G.-Munos, J. R.-Carvajel and P. Lacorre, Europhys. Lett. 20, 241 (1992).

[42] J. Rodríguez-Carvajal, S. Rosenkranz, M. Medarde, P. Lacorre, M. T. Fernandez-Díaz, F. Fauth, and V. Trounov, Phys. Rev. B 57, 456 (1998).

**[43]** Y. Tomioka, A. Asamitsu, H. Kuwahara, Y. Moritomo, and Y. Tokura, Phys. Rev. B **53**, R1689(1996).
.

[44] H. Yoshizawa, H. Kawano, Y. Tomioka and Y. Tokura, Phys. Rev. B52, R13145 (1995).

[45] Li X G, Zheng R K, Li G, Zhou H D and Huang R X, Europhys Lett. 60 (2002).

[46] R. Hamdi, A. Tozri, M. Smari, E. Dhahri and L. Bessais, ,Mat. Res. Bullet. 95, 525 (2017) .

**[47]** P. G. Radaelli, D. E. Cox, M. Marezio, and S-W. Cheong Phys. Rev. B **55**, 3015 **(1997).**

**[48]** S. Mori, C. H. Chen, and S-W. Cheong, Phys. Rev. Lett. **81**, 3972 (1998).